\documentstyle[12pt,twoside]{article}
\input{epsf}
\begin{document}
\setlength{\baselineskip}{12pt}
\hoffset 0.65cm
\voffset 0.3cm
\bibliographystyle{normal}

\vspace{48pt}
\noindent
{\bf Dynamics of Nematic Liquid Crystal under Oscillatory Flow: 
Influence of Surface Viscosity}

\vspace{48pt}
\noindent
I.Sh. NASIBULLAYEV, A.P. KREKHOV and M.V. KHAZIMULLIN

\noindent
Institute of Molecule and Crystal Physics, Russian Academy of Sciences, 450025 Ufa, Russia

\vspace{36pt}
\noindent
We analyse the influence of a surface viscosity on the orientational 
dynamics of a nematic liquid crystal subjected to an oscillatory
Couette flow.
Approximate analytical solutions of nematohydrodynamic equations for 
small flow amplitudes are calculated and compared with the results
of full numerical simulations.
The range of flow frequencies where the surface viscosity has strong 
influence on the optical response is determined.

\vspace{24pt}

\vspace{36pt}
\baselineskip=1.2 \baselineskip

\noindent
{\bf INTRODUCTION}
\vspace{12pt}

\noindent
In a nematic liquid crystal (NLC) complex flow behavior arises 
from the strong coupling between the velocity and orientational
degrees of freedom (director ${\bf \hat n}$).
Orientational dynamics of NLC under oscillatory flow
has generally been studied in the case of strong surface anchoring
(fixed orientation at the substrates).
The influence of weak surface anchoring and surface
viscosity on the dynamics of director reorientations under an applied
electric field has been discussed in \cite{GFD92,KL98}.
The surface-dominated orientational dynamics in NLC embedded in a 
solid porous matrix has been studied by means of the dynamic scattering 
technique, from which the temperature behavior of surface viscosity has 
been deduced \cite{MC98}.
Recently, a flow-induced orientational transition in nematics at the 
substrate with weak planar anchoring was found and an estimate of
surface viscosity has been given \cite{KBKL99}.

In this paper the orientational behavior of the NLC layer 
with weak surface anchoring subjected to the rectilinear oscillatory 
Couette flow is studied theoretically.
We analyse the case of small flow amplitudes where the director 
motion is within the flow plane.
The influence of anchoring strength and surface viscosity on the
orientational dynamics are investigated and the appropriate experimental
conditions to measure the surface friction are proposed.

\vspace{24pt}
\noindent
{\bf BASIC EQUATIONS}
\vspace{12pt}

\noindent
The nematic layer of thickness $d$ is confined between two identical substrates
which provide weak surface anchoring at $z=\pm d/2$ in a Cartesian coordinate system.
The oscillating flow is in the $x$ direction and the director confined to 
$x-z$ plane.
In this situation the director and the velocity are only functions 
of the distance $z$ from the boundaries and time $t$
\begin{eqnarray}
\label{eqn_1}
&n_x=\cos\theta(z,t),\;n_y=0,\;n_z=\sin\theta(z,t),&\nonumber\\
&v_x=v_x(z,t),\;v_y=0,\;v_z=0,&
\end{eqnarray}
where $\theta(z,t)$ is the angle with respect to the $x$ axis.
With the dimensionless variables $\tilde z=z/d$, $\tilde t=t\omega$, 
$\tilde v_x=v_x/(\omega d)$ the equations governing the alignment and the flow
can be rewritten as \cite{dGP93,KK96}
\begin{eqnarray}
\label{eqn_2}
&&\theta_{,t}-K(\theta)v_{x,z}=\varepsilon\left[P(\theta)\theta_{,zz}+
\frac{1}{2}P'(\theta)\theta^2_{,z}\right],\\
\label{eqn_3}
&&\delta v_{x,t} = \partial_z\Bigl\{-(1-\lambda)K(\theta)\theta_{,t}+
Q(\theta)v_{x,z}\Bigr\},
\end{eqnarray}
where the tildes have been omitted and
\begin{eqnarray*}
K(\theta)=\frac{\lambda\cos^2\theta-\sin^2\theta}{1-\lambda} ,
\lambda=\frac{\alpha_3}{\alpha_2} ,
P(\theta)=\cos^2\theta+K\sin^2\theta ,
K=\frac{K_{33}}{K_{11}} ,\\
2(-\alpha_2)Q(\theta)=
\alpha_4+(\alpha_5-\alpha_2)\sin^2\theta+
(\alpha_3+\alpha_6+2\alpha_1\sin^2\theta)\cos^2\theta ,\;\; \\
\varepsilon=\frac{1}{\tau_d \omega} ,
\tau_d=\frac{\gamma_1 d^2}{K_{11}} ,
\gamma_1=\alpha_3-\alpha_2 ,
\delta=\tau_v \omega ,
\tau_v=\frac{\rho d^2}{-\alpha_2} . \qquad \qquad
\end{eqnarray*}
Here the $\alpha_i$ are the viscosity coefficients, $K_{ii}$ are elastic constants, $\rho$
is the density of NLC and the notation $f_{,i} \equiv \partial f/\partial i$, 
$f'(g) \equiv \partial f/\partial g$ has been used throughout.
Boundary conditions for the velocity for the oscillatory Couette flow are
\begin{eqnarray}
\label{eqn_4}
&v_x(z=-1/2)=0,\;v_x(z=+1/2)=a\cos(t),&
\end{eqnarray}
where $a=A_x/d$ and $A_x$ is the upper plate displacement amplitude.

The weak anchoring is described mathematically in terms of surface energy
per unit area $F_s = (W/2) f_s(\theta-\theta_0)$,
where $W$ is the anchoring strength and function $f_s(\theta-\theta_0)$ 
has a minimum at $\theta=\theta_0$.
A simple phenomenological expression for the surface energy was introduced by
Rapini and Papoular where $f_s(\theta-\theta_0)=\sin^2(\theta-\theta_0)$ \cite{RP69}.
The boundary conditions for the director can be obtained from the surface torques
balance equation \cite{KL98,DV99}
\begin{eqnarray}
\label{eqn_6}
&&\mp P(\theta)\theta_{,z}+\frac{W d}{2 K_{11}}\frac{\partial f_s}{\partial\theta}
+\frac{\omega d}{K_{11}}\eta\;\frac{\partial\theta}{\partial t}=0,\\
&&- \; \mbox{on} \;\; z=-1/2 \;\;\; \mbox{and} + \mbox{on} \;\; z=1/2 .\nonumber
\end{eqnarray}
Here $\eta=\gamma_1 l_{\gamma_1}$ is so-called surface viscosity which characterises the
dissipation at the substrate; $\gamma_1$ is the
bulk orientational viscosity and $\l_{\gamma_1}$ 
is a characteristic interfacial length for surface viscosity.

In order to obtain an approximate analytical solution of 
Eqs.(\ref{eqn_2}), (\ref{eqn_3}) we consider
the case of small flow amplitudes $a \ll 1$ which corresponds to the small distortions
of the director profile
\begin{equation}
\label{eqn_7}
\theta=\theta_0 + \tilde\theta \; , \;\; v_x=v_{x0} + U \; , 
\;\; |\tilde\theta| \ll 1\; , \;\;  |U| \ll 1 \; ,
\end{equation}
where $\theta_0=const$, $v_{x0}=0$ is the solution of (\ref{eqn_2}), (\ref{eqn_3})
in the absence of oscillatory flow ($a=0$); the value $\theta_0$ is defined
by the minimum of surface energy given 
for small amplitude of the director oscillations as 
$f_s=\tilde\theta^2$.
In the low-frequency range to be considered here one has
$\delta \ll 1$ ($\rho \approx 10^3$ kg/m$^3$, $|\alpha_2| \approx 10^{-1}$
N$\cdot$s/m$^2$ and $d \approx 10^{-5}$ m gives $\delta < 1$ for frequencies
$f<1$ kHz) and the inertia term [left-hand side of 
Eq.(\ref{eqn_3})] is neglected.
Then one obtains from equations (\ref{eqn_2}), (\ref{eqn_3}) 
\begin{eqnarray}
\label{eqn_8}
&&\tilde\theta_{,t}-K_0 U_{,z}=\varepsilon P_0 \tilde\theta_{,zz} , \nonumber\\
&&(1-\lambda)K_0\tilde\theta_{,tz}-Q_0 U_{,zz}=0 ,
\end{eqnarray}
where $K_0=K(\theta_0)$, $P_0=P(\theta_0)$, $Q_0=Q(\theta_0)$, with the
boundary conditions
\begin{eqnarray}
\label{eqn_9}
&&\tilde\theta_{,z}-E \tilde\theta - G \tilde\theta_{,t} = 0 \mid_{z=-1/2} \; , \;\; 
\tilde\theta_{,z}+E \tilde\theta + G \tilde\theta_{,t} = 0 \mid_{z=1/2}, \nonumber \\
&&U(z=-1/2) = 0, \;\; U(z=1/2) = a\cos(t).
\end{eqnarray}
Here $E=W d/(P_0 K_{11})$, $G=\omega d \gamma_1 l_{\gamma_1}/(P_0 K_{11})$.
Since Eqs.(\ref{eqn_8}) are linear, the periodic
boundary conditions (\ref{eqn_9}) (periodic forcing) will lead to time-periodic
solutions for the director
\begin{equation}
\label{eqn_10}
\tilde\theta(z,t) = T_1(z)\cos(t) + T_2(z)\sin(t)
\end{equation}
with
\begin{eqnarray*}
T_1(z)=a K_0\frac{C_1 F_1(z)-C_2 F_2(z)}{C_1^2+C_2^2}, \;\;
T_2(z)=-a K_0\frac{C_1 F_2(z)+C_2 F_1(z)}{C_1^2+C_2^2}
\end{eqnarray*}
%
%
and for the velocity
\begin{eqnarray}
\label{eqn_12}
&&U(z,t) = U_1(z)\cos(t)+U_2(z)\sin(t), \\
&&U_1(z)=a(\frac{1}{2}+z)+a\frac{(1-\lambda)K_0^2}{2 Q_0 k}
        \frac{C_1 F_3(z)-C_2 F_4(z)}{C_1^2+C_2^2},\nonumber\\
&&U_2(z)=-a\frac{(1-\lambda)K_0^2}{2 Q_0 k}
\frac{C_1 F_4(z)+C_2 F_3(z)}{C_1^2+C_2^2} \; , \nonumber
\end{eqnarray}
where the functions $F_i(z)$ and $C_i$ are given in Appendix A.

\vspace{24pt}
\noindent
{\bf RESULTS AND DISCUSSION}
\vspace{12pt}

\noindent
For weak homeotropic anchoring 
($\theta_0=\pi/2$, $W=10^{-6}$ J/m$^2$)
the profiles of the director
deviation $T_1(z)$, $T_2(z)$
from the homeotropic position as well as the velocity components
$U_1(z)$, $U_2(z)$ are presented in Fig.~1 for different values of
the surface viscosity $\eta=\gamma_1 l_{\gamma_1}$.
The material parameters of MBBA  (see Appendix B) have been used
throughout.
The curves for strong homeotropic anchoring [fixed orientation at the
substrates, $\tilde\theta(z=\pm 1/2)=0$] as well as for zero surface 
viscosity ($l_{\gamma_1}=0$) are also shown for comparison.
For the velocity component $U_1$ we subtract the linear part $a(z+1/2)$ which corresponds
to the velocity profile in case of isotropic liquid.
One can see that the value of surface viscosity strongly influence on the director and velocity
distribution under the oscillatory Couette flow.
Note, that raising of the surface viscosity has a similar effect as the
increasing of the surface anchoring strength $W$.

\begin{figure}
\begin{center}
\vspace*{-0.1cm}
\hspace*{-1cm}
\epsfxsize=10.5cm
\epsfbox{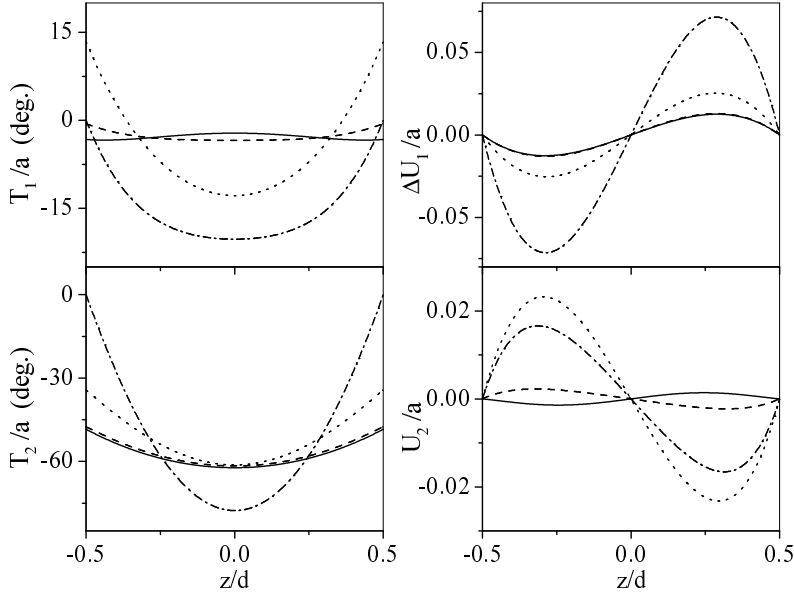}
\end{center}
\vspace*{-1.2cm}
\caption{Director and velocity profiles for oscillatory Couette flow.
$a=0.2$, $f=5$ Hz, $d=10$ $\mu$m, $W=10^{-6}$ J/m$^2$,
$l_{\gamma_1}$ [m]: $0$ (---); $10^{-7}$ (-- --); $10^{-6}$ ($\cdots$), 
strong anchoring (-- $\cdot$ --).}
\label{fig1}
\end{figure}

The amplitude of the director oscillations is proportional to 
$K_0=K(\theta_0)$.
For MBBA one has $K(\pi/2)/K(0) \approx 10^2$ and the amplitude of the director
oscillations for weak planar anchoring is much smaller than for the homeotropic
orientation at the same flow amplitude $a$.
Therefore, in case of planar orientation the influence of the surface viscosity
on the orientational dynamic of NLC is much smaller than for weak homeotropic
anchoring.

One of the widely used technique for studying the orientational behavior
of NLC is the
measurements of the transmitted through the NLC cell polarized light.
In the geometry where the polars are crossed and the $x$ axis is at 
$45^{\circ}$ one has for the light intensity
\begin{equation}
\label{eqn_20}
I = I_0 \sin^2\frac{\Psi}{2} \; , \;\;
\Psi = \frac{2\pi d}{\Lambda}\int \limits_{-1/2}^{1/2}
\Bigg[ 
\frac{n_o n_e}{\sqrt{n_o^2\cos^2\theta + n_e^2\sin^2\theta}} - n_o 
\Bigg] d z \; ,
\end{equation}
where $n_o$, $n_e$ are the ordinary and extraordinary refractive 
indices and $\Lambda$ is
the wavelength of light.
Time-periodic director oscillations lead to a variation of the optical
response $I=I(t)$.
Using the director solution (\ref{eqn_10}), the transmitted 
light intensity can easily be calculated.
\begin{figure}[t]
\begin{center}
\vspace*{-0.2cm}
\hspace*{-1.5cm}
\epsfxsize=9.0cm
\epsfbox{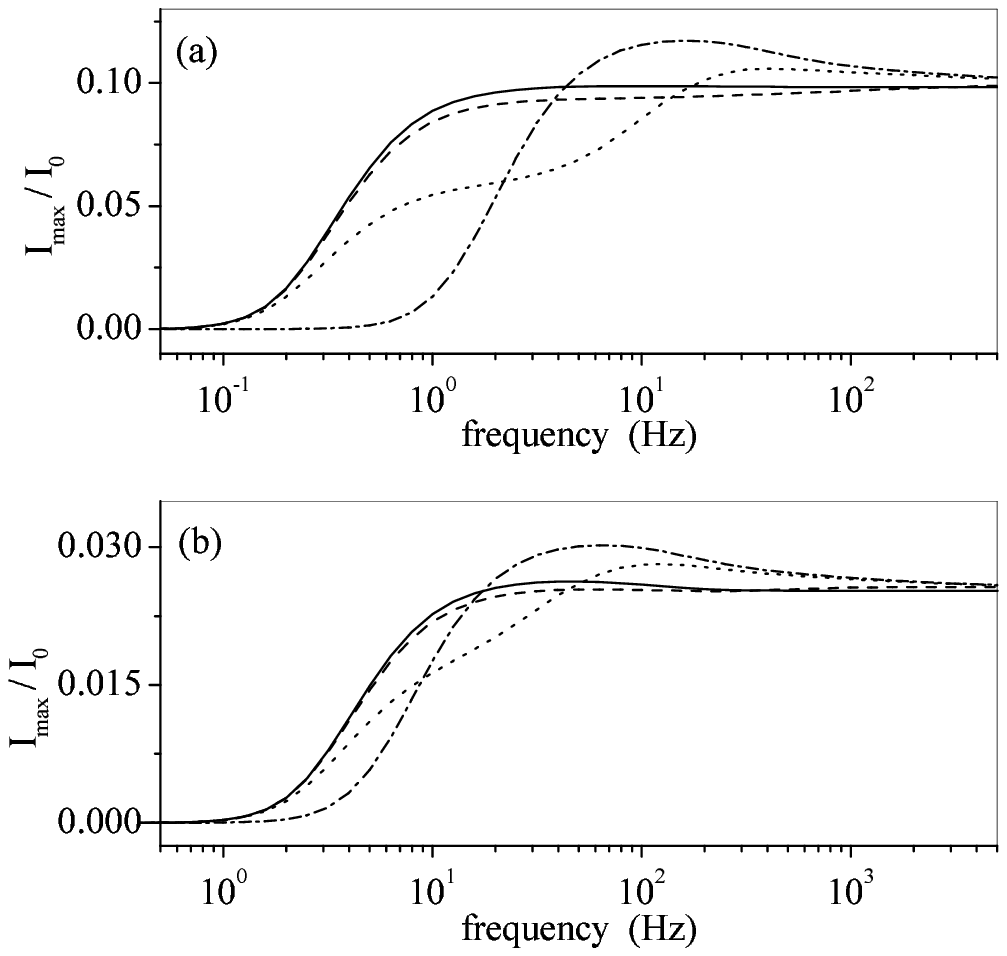}
\end{center}
\vspace*{-1.2cm}
\caption{Maximum of the transmitted light intensity versus the 
frequency of oscillatory Couette flow at $a=0.2$.
(a) - $d=10$ $\mu$m, $W=10^{-6}$ J/m$^2$;
(b) - $d=5$ $\mu$m, $W=10^{-5}$ J/m$^2$.
$l_{\gamma_1}$ [m]: $0$ (---); $10^{-7}$ (-- --); $10^{-6}$ ($\cdots$),
strong anchoring (-- $\cdot$ --).}
\label{fig2}
\end{figure}
We found that depending on the value of surface viscosity, the maximum 
of the intensity changes and its position is shifted in time with
respect to the moment of zero displacement of upper plate ($t=\pi/2$).
In order to find the range of flow frequencies where the surface viscosity has a strong influence
on the optical response, the dependence of the maximum of the transmitted light intensity
$I_{max}$ on the flow frequency was calculated for different values of surface viscosity
(Fig.~2).
We present the results for two typical values of anchoring strength $W$
within the range of recently found experimental ones \cite{MC98}.
The optical response for the case of strong anchoring is shown for
comparison.
It is clearly seen that there exists some flow frequency
range [$\sim 0.1 \div 10$ Hz in Fig.~2(a) and
$\sim 1 \div 100$ Hz in Fig.~2(b)] where the optical response
is strongly depending on the value of surface viscosity.
Below this frequency range the transmitted intensity is small and
the determination of the surface viscosity is complicated, whereas at 
higher frequencies the boundary layers of the order of 
$\sqrt{K_{11}/(2\omega\gamma_1)}$ at $z=\pm 1/2$ \cite{KK96}
become thinner and the influence of the surface viscosity on the 
director dynamics is not so apparent.

In order to verify the approximate solutions, direct numerical simulations
of system (\ref{eqn_2}), (\ref{eqn_3}) with boundary conditions 
(\ref{eqn_4}), (\ref{eqn_6}) have been performed.
It was found that for the frequency range $\omega \ll \tau_v^{-1}$ 
($\delta \ll 1$) one can safely
use the analytical small-amplitude solutions up to the flow
amplitudes $a \approx 0.25$ for the homeotropic anchoring (the difference
between analytical expressions and numerical solution does not exceed $0.5\%$).

\vspace{12pt}
In conclusion, the influence of the anchoring strength and surface 
viscosity on the
director dynamics in the bulk of NLC and at the substrates is 
investigated for oscillatory Couette flow.
Approximate analytical solutions for small-amplitude oscillatory 
flow in case of weak surface anchoring are obtained.
The analysis shows that there exists a certain flow frequency range 
where the 
optical response strongly depends on the value of surface viscosity.
This effect can be used for a more precise experimental determination of the surface
viscosity and investigations of surface-dominated orientational
dynamics in oscillatory flow of NLC.

\vspace{12pt}
\noindent
{\bf Appendix A:}
\vspace{-8pt}
\noindent
%
\begin{eqnarray*}
&&F_1(z)=\cosh(kz)\cos(kz)-\cosh(k/2)\cos(k/2)+\\
        &&\qquad\qquad+\frac{kE}{E^2+G^2}\Bigl\{\cosh(k/2)\sin(k/2)-
        \sinh(k/2)\cos(k/2)\Bigr\}-\\
        &&\qquad\qquad-\frac{kG}{E^2+G^2}\Bigl\{\cosh(k/2)\sin(k/2)+
        \sinh(k/2)\cos(k/2)\Bigr\},\\
&&F_2(z)=\sinh(kz)\sin(kz)-\sinh(k/2)\sin(k/2)-\\
        &&\qquad\qquad-\frac{kE}{E^2+G^2}\Bigl\{\cosh(k/2)\sin(k/2)+
        \sinh(k/2)\cos(k/2)\Bigr\}-\\
        &&\qquad\qquad-\frac{kG}{E^2+G^2}\Bigl\{\cosh(k/2)\sin(k/2)-
        \sinh(k/2)\cos(k/2)\Bigr\},\\
&&F_3(z)=\sinh(kz)\cos(kz)-\cosh(kz)\sin(kz)-\\
        &&\qquad\qquad-2z\Bigl(\sinh(k/2)\cos(k/2)-\cosh(k/2)\sin(k/2)\Bigr),\\
&&F_4(z)=\sinh(kz)\cos(kz)+\cosh(kz)\sin(kz)-\\
        &&\qquad\qquad-2z\Bigl(\sinh(k/2)\cos(k/2)+\cosh(k/2)\sin(k/2)\Bigr), \\
&&k=\sqrt{\frac{Q_0-(1-\lambda)K_0^2}{2\varepsilon Q_0 P_0}}, \\
&&C_1=\sinh(k/2)\sin(k/2)-\\
        &&\qquad\qquad-\frac{(1-\lambda)K_0^2}{Q_0 k}\Bigl[\cosh(k/2)\sin(k/2)
        -\sinh(k/2)\cos(k/2)\Bigr]+\\
        &&\qquad\qquad+\frac{kE}{E^2+G^2}\Bigl\{\cosh(k/2)\sin(k/2)+
        \sinh(k/2)\cos(k/2)\Bigr\}+\\
        &&\qquad\qquad+\frac{kG}{E^2+G^2}\Bigl\{\cosh(k/2)\sin(k/2)-
        \sinh(k/2)\cos(k/2)\Bigr\},\\
&&C_2=\cosh(k/2)\cos(k/2)-\\
        &&\qquad\qquad-\frac{(1-\lambda)K_0^2}{Q_0 k}\Bigl[\cosh(k/2)\sin(k/2)
        +\sinh(k/2)\cos(k/2)\Bigr]-\\
        &&\qquad\qquad-\frac{kE}{E^2+G^2}\Bigl\{\cosh(k/2)\sin(k/2)-
        \sinh(k/2)\cos(k/2)\Bigr\}+\\
        &&\qquad\qquad+\frac{kG}{E^2+G^2}\Bigl\{\cosh(k/2)\sin(k/2)+
        \sinh(k/2)\cos(k/2)\Bigr\} \; .
\end{eqnarray*}

\noindent
{\bf Appendix B:}

\noindent
The numerical computations are carried out for the following MBBA
material parameters at 
$25$ $^{\circ}$C \cite{dJCS76,KSS82}:
viscosity coefficients in units of $10^{-3}$ N$\cdot$s/m$^2$ :
$\alpha_1=-18.1$, $\alpha_2=-110.4$, $\alpha_3=-1.1$,
$\alpha_4=82.6$, $\alpha_5=77.9$, $\alpha_6=-33.6$;
elasticity coefficients in units of $10^{-12}$ N:
$K_{11}=6.66$, $K_{22}=4.2$, $K_{33}=8.61$;
mass density $\rho=10^3$ kg/m$^3$
and refractive indices for wavelength of light $\Lambda=670$ nm: 
$n_o=1.542$, $n_e=1.7435$ \cite{B70}.

\vspace{12pt}
\noindent
{\bf Acknowledgments}

\noindent
We thank W. Pesch and Yu. Lebedev for fruitful discussions and 
critical reading of the manuscript.
Financial support from INTAS (Grant No. 96-498) and DFG (Grant No. 
Kr690/12-1) is gratefully acknowledged.

\setlength{\baselineskip}{12pt}

\end{document}